\documentclass[manuscript]{acmart}
\usepackage[braket, qm]{qcircuit}
\usepackage{bm}
\usepackage{listings}

\setcopyright{none}

\DeclareMathOperator*{\argmin}{arg\,min}
\usepackage[caption=false]{subfig}
\usepackage[singlelinecheck=false]{caption} 

\usepackage{hyperref}
\usepackage{graphicx}
\usepackage{comment}
\usepackage{amsmath}
\usepackage{bbm}
\usepackage{algorithm}
\usepackage{algpseudocodex}
\usepackage{braket}
\usepackage{verbatim}
\usepackage{tabularx}
\usepackage{cite}
\usepackage{natbib}
\usepackage{color, colortbl}
\usepackage{multirow}
\definecolor{LightCyan}{rgb}{1,0.5,0.5}
\hypersetup{
  colorlinks   = true, 
  urlcolor     = blue, 
  linkcolor    = blue, 
  citecolor   = blue 
}

\begin{document}
%
\title{\texttt{QArchSearch}: A Scalable Quantum  Architecture Search Package}


\author{Ankit Kulshrestha}
\affiliation{%
 \institution{University of Delaware}
 \city{Newark}
 \state{DE}
 \country{USA}}

\author{Danylo Lykov}
\affiliation{%
 \institution{Argonne National Laboratory}
 \city{Lemont}
 \state{IL}
 \country{USA}}

\author{Ilya Safro}
\affiliation{%
 \institution{University of Delaware}
 \city{Newark}
 \state{DE}
 \country{USA}}

\author{Yuri Alexeev}
\affiliation{%
 \institution{Argonne National Laboratory}
 \city{Lemont}
 \state{IL}
 \country{USA}}

\begin{abstract}
The current era of quantum computing has yielded several algorithms that promise high computational efficiency. While the algorithms are sound in theory and can provide potentially exponential speedup, there is little guidance on how to design proper quantum circuits to realize the appropriate unitary transformation to be applied to the input quantum state. In this paper, we present \texttt{QArchSearch}, an AI based quantum architecture search package with the \texttt{QTensor} library as a backend that provides a principled and automated approach to finding the best model given a task and input quantum state. We show that the search package is able to efficiently scale the search to large quantum circuits and enables the exploration of more complex models for different quantum applications. \texttt{QArchSearch} runs at scale and high efficiency on high-performance computing systems using a two-level parallelization scheme on both CPUs and GPUs, which has been demonstrated on the Polaris supercomputer.
\end{abstract}


\maketitle


%

\section{Introduction}
%
%
%
%

 Quantum computing is a nascent and rapidly growing field that holds the promise of accomplishing tasks that were hitherto thought too be computationally intractable by classical computers. In the current era, we have access to noisy intermediate scale quantum computers (NISQ) that make it possible to run hybrid quantum algorithms to tackle problems in computational chemistry, finance \citep{herman2023quantum}, optimization \citep{ushijima2021multilevel} and related fields \citep{shaydulin2019network}. These algorithms leverage a ``variational'' method in which quantum circuit parameters have to be trained through a classical optimization procedure (generally run on a classical co-processor).

 In their most abstract form, variational quantum circuits are a form of linear parameterized unitary transformation $U(\bm{\theta})$ that map an input quantum state $\ket{\psi}_{in}$ to an output quantum state $\ket{\psi}_{out} = U(\bm{\theta})\ket{
\psi}_{in}$. Currently, the goal in variational quantum algorithms (VQAs) is to find $\bm{\theta}^* = \argmin_{\theta} C(\bm{\theta})$ where $C(\bm{\theta})$ is some cost function that quantifies the quality of output. However, in this paper we consider an alternative problem: We aim to find the best possible circuit representing $U(\bm{\theta})$ given input $\ket{\psi}_{in}$ and cost function $C(\bm{\theta})$.

Finding an appropriate quantum circuit architecture for a given application is a computationally intensive search procedure that requires evaluation of  several candidate quantum operations across different qubits and selecting the best performing circuit from amongst them. Our focus in this paper is to demonstrate how we scale such a search procedure across a state of the art HPC infrastructure to aid the search by training deep neural networks to suggest good circuit structures. We title our software as \texttt{QArchSearch} and include it as a part of widely available \texttt{QTensor} package.


\begin{figure}
    \centering
    \includegraphics[width=\columnwidth]{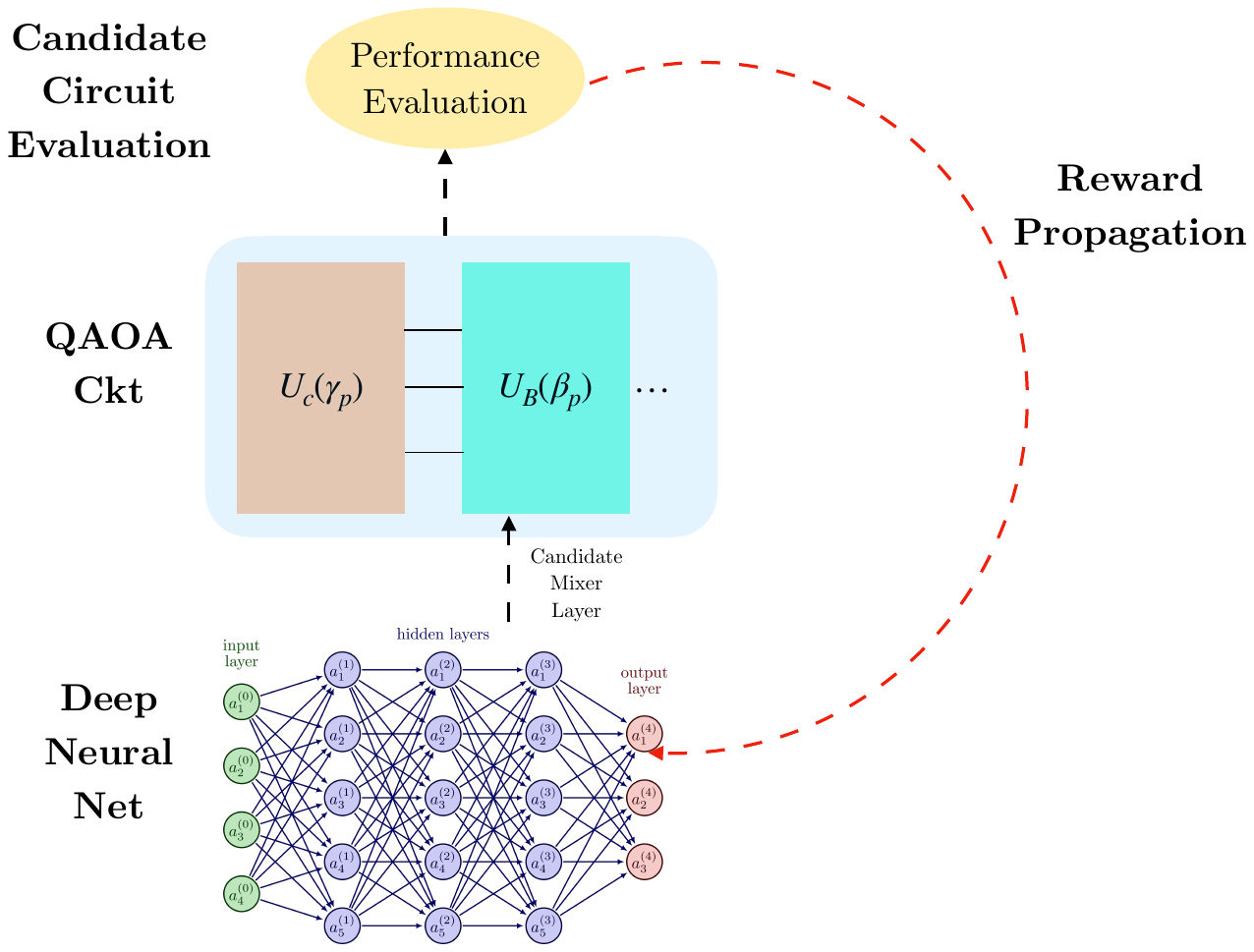}
    \caption{An overview of the search process in \texttt{QArchSearch} software}
    \label{fig:qarchsearch_flow}
\end{figure}

In this work, we will use the Quantum Approximate Optimization Algorithm (QAOA)~\citep{farhi2014quantum} for the graph maxcut problem as the driver application of the \texttt{QArchSearch} package. Briefly, for a given simple undirected graph $G=(\mathcal{V}, \mathcal{E})$, the graph maxcut problem aims to find a maximum ``cut set'', i.e., a partition of nodes into two disjoint parts to maximize the number (or the total weight in case the graph is weighted) of edges that span both parts. The cost function for max cut is given as~\citep{farhi2014qaoa_for_bounded}:
\begin{equation}\label{eq:maxcut_objective}
    C_{MC}(\bm{z}) = \frac{1}{2} \sum_{(u,v)\in \mathcal{E}} ( 1 - z_uz_v),
\end{equation}
where $\bm{z}_i \in \{-1, +1\}$ is an indicator variable for node $i$ that corresponds to the set membership the given node. In the QAOA setup, we start with an initial state $\ket{s} = \ket{+}^{\otimes^n}$ where $\ket{+} = \frac{\ket{0} + \ket{1}}{\sqrt{2}}$. A $p$ layer alternating ansatz is then applied to the initial input state:
\begin{equation}\label{eq:qaoa_ansatz}
    \ket{\bm{\gamma}, \bm{\beta}} = e^{-i\beta_p B} e^{-i\gamma_pC} \dots e^{-i \beta_1 B} e^{-i\gamma_1 C} \ket{s}.
\end{equation}

Here, $\bm{\gamma}, \bm{\beta} \in \mathbb{R}^p$ are parameters of the cost operator $C$ and the mixer operator $B$, respectively. The cost function is measured by computing $\bra{\bm{\gamma}, \bm{\beta}}C(\bm{z})\ket{\bm{\gamma}, \bm{\beta}}$. In QAOA problems, the structure of the cost operator $C$ is generally guided by the problem we are interested in optimizing, but the structure of mixer operator is an open design problem. In our application, \texttt{QArchSearch} is responsible for searching low-depth mixers  using the process depicted in Figure~\ref{fig:qarchsearch_flow}.



\section{Methodology}

\subsection{QArchSearch}
The \texttt{QArchSearch} software has three key components:
\begin{itemize}
    \item Predictor module: This module accepts a tensor that represents the rotation gates and entanglement operators and generates a new circuit representation that is passed to the quantum builder module.

    \item Quantum Builder (a.k.a QBuilder): This module accepts the encoded tensor representation from the predictor module and generates the appropriate quantum circuit in an available quantum computing software. In our work, the circuits are generated using \texttt{Qiskit} software. The generated circuit is then passed to the evaluator.

    \item Evaluator Module: This module is responsible for training the generated quantum circuit on the QAOA cost function in Equation~\ref{eq:maxcut_objective}. The trained circuit is then evaluated and the reward is propagated back to the predictor module. 
\end{itemize}

Algorithm~\ref{alg:qarchsearch} shows the overall search procedure for searching best performing QAOA mixer circuit from a given gate alphabet $\mathcal{A}_{R}$. The current version of the search algorithm is an instance of random search which has shown to be a strong baseline in neural architecture search~\citep{li2020random}. We perform a search by varying the depth $p$ from 1 to desired maximum depth. For each $p$ we explore the best possible gate combination (Line 5) and construct a mixer circuit based on the nodes in the current graph (Line 6). We then instantiate the QAOA ansatz and run the variational algorithm for 200 steps with the COBYLA optimizer. The obtained energy is added to a global collection (Line 9). At the end of exploring all possible gate combinations, we select the best performing mixer circuit and compare it to the previously existing best performing mixer circuit if it exists (Line 10). The final best performing mixer circuit and corresponding cut energy are then returned to the main calling procedure. 
 
\begin{algorithm}[h]
\caption{\texttt{QArchSearch} for QAOA Mixer} \label{alg:qarchsearch}
    \begin{algorithmic}[1]
        \Require $\bm{\theta}$: parameters of quantum circuit, $\mathcal{A}_{R}$: Gate alphabet, $p_{max}$: depth of QAOA ansatz, $K_{max}$: maximum number of possible gate combinations
        $G$: input graph

        \State $U^{best}_{B} \gets \phi$ 
        \For{$p: 1 \dots p_{max}$ }
            \State energies $\gets$ $\{\}$
            \For{$k: 1 \dots K_{max}$}
                
                \State gate\_comb $\gets$ \Call{GET\_COMBINATIONS}{$\mathcal{A}_{R}$, k}
                
                \State $\hat{U}_B \gets$ \Call{BUILD\_MIXER\_CKT}{$G$, gate\_comb}
                \State $U_{QAOA}(\bm{\theta}) \gets$ \Call{BUILD\_QAOA\_CKT}{$\hat{U}_B, p$}
                \State $\langle C \rangle \gets$ \Call{SIMULATE\_QAOA}{$G, U_{QAOA}(\bm{\theta}$}
                \State energies $\gets $ \Call{APPEND}{energies, $\langle C \rangle$}
                \EndFor
            \State $U^{best}_{B} \gets $ \Call{SELECT\_BEST}{energies, $U^{best}_{B}$}
            
         \EndFor
         \Return \State $U^{best}_B$, $\langle C_{best} \rangle$

    \end{algorithmic}
\end{algorithm}

\begin{figure}
    \centering    
    \includegraphics[width=0.65\linewidth]{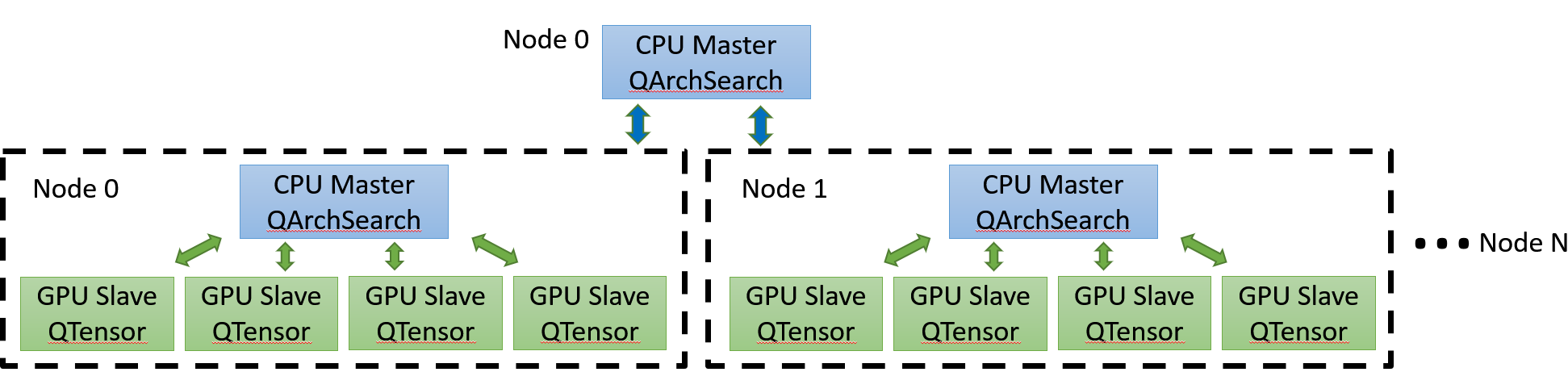}
    \caption{
    The parallelization architecture within each node and between nodes using the \texttt{QArchSearch} and \texttt{QTensor} packages on the Polaris supercomputer. 
    }
\label{fig:architecture}
\end{figure}

\subsection{QTensor}
In this work, we used the Argonne-developed tensor network simulator~\citep{lykov_step_dependent_qtensor,LykovSampling2023,Lykov_diagonal_gates}. It is developed for running large-scale quantum circuit simulations using modern GPU-based supercomputers. It has been used to perform the largest QAOA simulations in the world. \texttt{QTensor} utilizes state-of-the-art heuristic tensor contraction order optimizers (third-party and own custom optimizers), which substantially reduce the simulation cost by minimizing the contraction width of the contraction sequence. We used a number of techniques to speed up simulations.

\texttt{QTensor} has support for a few tensor contraction libraries (backends) for contracting tensors efficiently. In this work, we used NumPy for tensor contraction on CPUs. The code is freely available on GitHub~\citep{QTensor}.

\section{Experiments and Results}

In this section we present our results on the single and multi-core performance of \texttt{QArchSearch}. We then show that the circuit resulting from our search procedure generalizes to unseen graph instances and can achieve better max-cut energies even at low depths. 

\begin{figure}[h]
    \centering
    \includegraphics[width=0.65\linewidth]{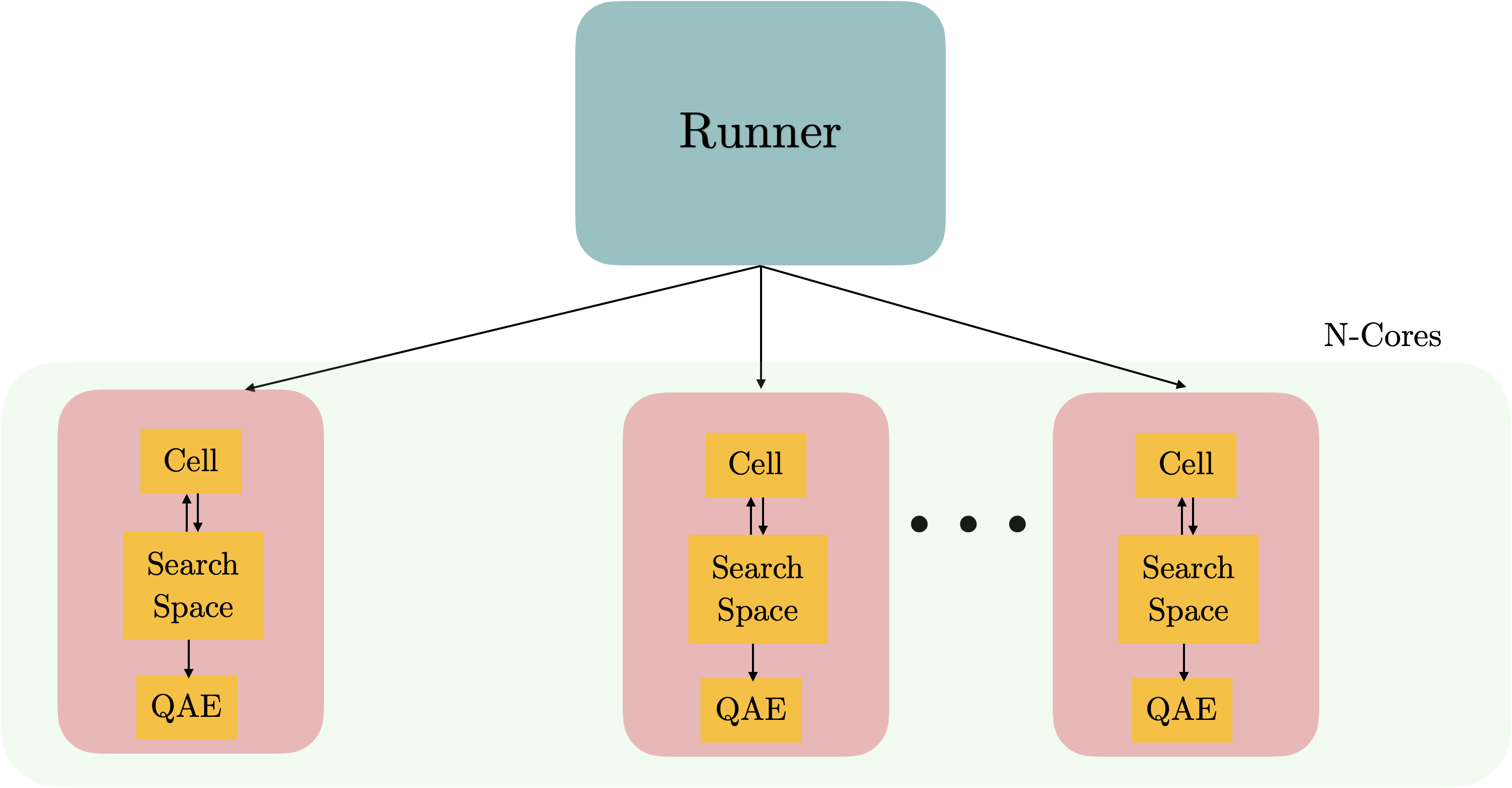}
    \caption{The CPU level parallelism exposed by current version of \texttt{QArchSearch}.}
    \label{fig:qarchsearch_current}
\end{figure}

\subsection{Performance Profiling Results}

To perform the performance profiling, we first implemented a serially executed search procedure that examined every possible rotation gate combination and simulated the resulting circuit for depths $p=1 \dots 4$. For each depth, we performed a combination of $k=1\dots 4$ gates over the given rotation gate alphabet $\mathcal{A}_{R}$ with $|\mathcal{A}_{R}| = 5$ leading to 2500 possible circuit combinations. All search profiling was performed on a dataset of 20, 10-node Erdos-Renyi graphs with varying degrees of connectivity.

\textbf{Serial Search Process}: We first profiled a serial search process that sequentially examined each possible gate combination for a given depth $p$. The expected run time of the algorithm was thus $O(pk)$ where $k$ was the number of gates selected from $\mathcal{A}_{R}$.\\

\begin{figure}[h]
    \centering
    \includegraphics[width=.65\linewidth]{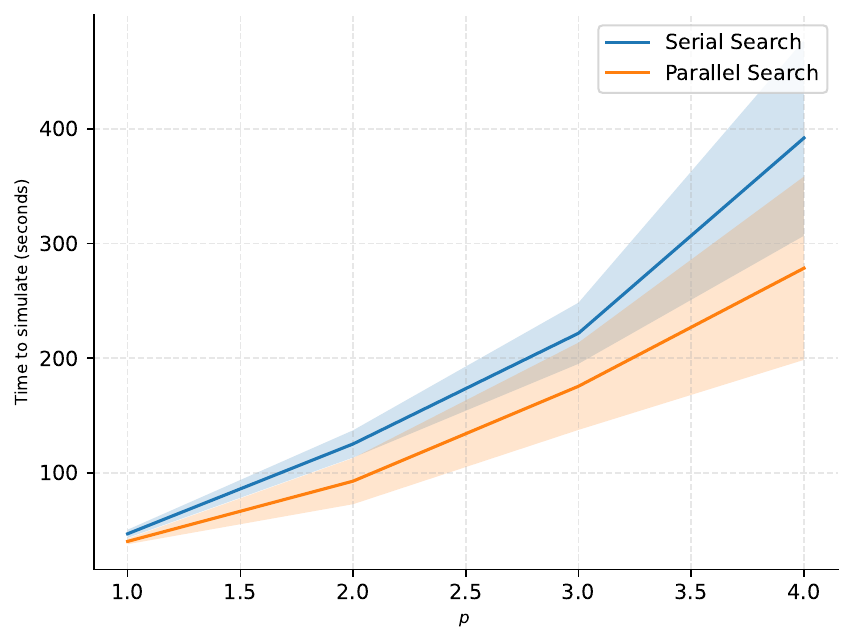}
    \caption{Time to simulate circuits with serial and parallel quantum NAS procedure. The results are averaged over five separate runs of the NAS algorithm on different Erdos-Renyi Graphs}
    \label{fig:ser_vs_par}
\end{figure}

\textbf{Parallelizing Architecture Search}: To speedup the search process, it was necessary to parallelize the algorithm without causing a degradation in the quality of search results. We identified the sequential simulation of gate combinations for a given graph and depth as a major computational bottleneck. Hence, our focus was to improve run time by searching multiple possible gate combinations in parallel. This strategy is shown in Figure~\ref{fig:qarchsearch_current}\\ 

To accomplish the aforementioned objective, we opted for process-level parallelism that can take advantage of multiple CPUs on a single node of a HPC cluster. We used Python's \texttt{multiprocessing} library's \texttt{starmap\_async} method to create a pool of processes on different CPUs that executed the optimization objective of Equation~\ref{eq:maxcut_objective} with different gate combinations in parallel. The run time was thus reduced from $O(pk)$ to $O(p)$ for a single graph. \\


\begin{figure}[h]
    \centering
    \includegraphics[width=0.65\linewidth]{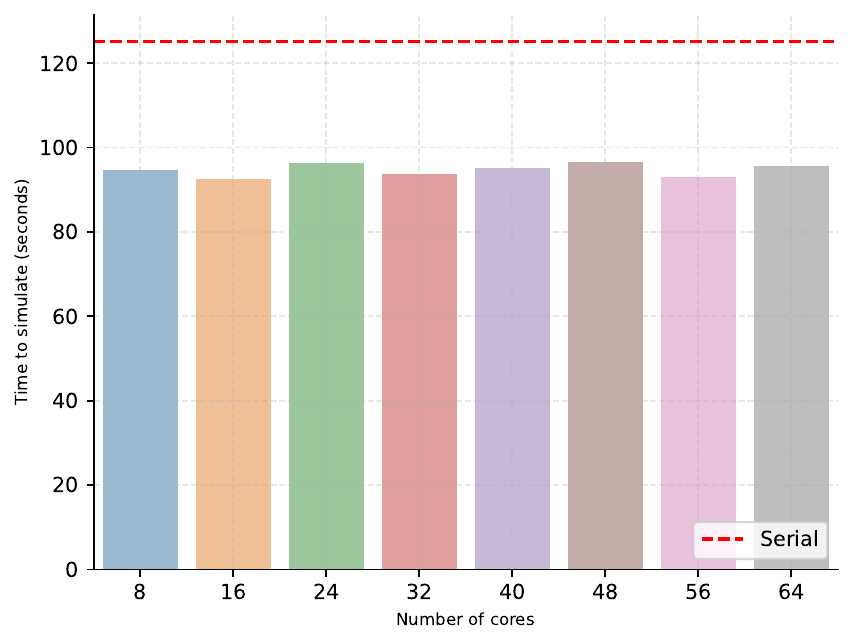}
    \caption{Time to simulate a graph with $p=2$ with different number of cores available on a HPC cluster. The dashed red line indicates the time to simulate the same graph with serial search.}
    \label{fig:core_perf}
\end{figure}

\textbf{Results}: The results of our profiling experiments are shown in Figure~\ref{fig:ser_vs_par} and Figure~\ref{fig:core_perf}. 
Figure~\ref{fig:ser_vs_par} shows the improvement in the run time of the algorithm with increasing depth of the QAOA ansatz. We note that in the serial case, the growth in the run time is quadratic as $p \approx k$. However, in the case of parallel the run time is improved by over 50\% even when the depth approaches the maximum possible gate combinations.

Figure~\ref{fig:core_perf} shows the time to simulate for a graph with $p=2$ and the number of available CPUs varied from 8 to 64 in increments of 8. We can see that our parallel version can efficiently utilize the available CPUs and are 0.76 times faster than a serial algorithm for the same graph and $p$.

\subsection{Performance of Discovered Circuit}

\begin{figure}[]
    \centering
    \scalebox{1.0}{
\Qcircuit @C=1.0em @R=0.2em @!R { \\
	 	\nghost{{q}_{0} :  } & \lstick{{q}_{0} :  } & \gate{\mathrm{R_X}\,(\mathrm{2\beta})} & \gate{\mathrm{R_Y}\,(\mathrm{2\beta})} & \qw & \qw\\
	 	\nghost{{q}_{1} :  } & \lstick{{q}_{1} :  } & \gate{\mathrm{R_X}\,(\mathrm{2\beta})} & \gate{\mathrm{R_Y}\,(\mathrm{2\beta})} & \qw & \qw\\
	 	\nghost{{q}_{2} :  } & \lstick{{q}_{2} :  } & \gate{\mathrm{R_X}\,(\mathrm{2\beta})} & \gate{\mathrm{R_Y}\,(\mathrm{2\beta})} & \qw & \qw\\
	 	\nghost{{q}_{3} :  } & \lstick{{q}_{3} :  } & \gate{\mathrm{R_X}\,(\mathrm{2\beta})} & \gate{\mathrm{R_Y}\,(\mathrm{2\beta})} & \qw & \qw\\
	 	\nghost{{q}_{4} :  } & \lstick{{q}_{4} :  } & \gate{\mathrm{R_X}\,(\mathrm{2\beta})} & \gate{\mathrm{R_Y}\,(\mathrm{2\beta})} & \qw & \qw\\
	 	\nghost{{q}_{5} :  } & \lstick{{q}_{5} :  } & \gate{\mathrm{R_X}\,(\mathrm{2\beta})} & \gate{\mathrm{R_Y}\,(\mathrm{2\beta})} & \qw & \qw\\
	 	\nghost{{q}_{6} :  } & \lstick{{q}_{6} :  } & \gate{\mathrm{R_X}\,(\mathrm{2\beta})} & \gate{\mathrm{R_Y}\,(\mathrm{2\beta})} & \qw & \qw\\
	 	\nghost{{q}_{7} :  } & \lstick{{q}_{7} :  } & \gate{\mathrm{R_X}\,(\mathrm{2\beta})} & \gate{\mathrm{R_Y}\,(\mathrm{2\beta})} & \qw & \qw\\
	 	\nghost{{q}_{8} :  } & \lstick{{q}_{8} :  } & \gate{\mathrm{R_X}\,(\mathrm{2\beta})} & \gate{\mathrm{R_Y}\,(\mathrm{2\beta})} & \qw & \qw\\
	 	\nghost{{q}_{9} :  } & \lstick{{q}_{9} :  } & \gate{\mathrm{R_X}\,(\mathrm{2\beta})} & \gate{\mathrm{R_Y}\,(\mathrm{2\beta})} & \qw & \qw\\ }}
    \caption{Best performing searched mixer circuit for Max-cut QAOA}
    \label{fig:searched_ckt}
\end{figure}
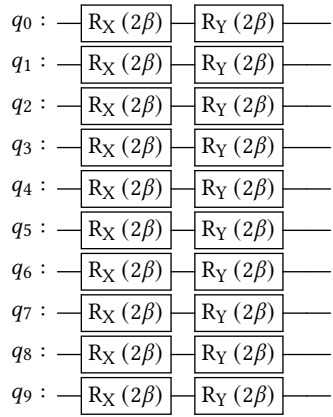

Once the search procedure was run, we evaluated the possible discovered combinations of the mixer layer on a separate dataset of 20, 10 node random 4-regular graphs. The best performing mixer circuit is shown in Figure~\ref{fig:searched_ckt}. For each discovered circuit, we calculated the approximation ratio $r$ defined as:

\begin{equation}\label{eq:approx_ratio}
    r = \frac{\langle C_{max} \rangle}{C_{classical}}
\end{equation}

\begin{figure}[h]
    \centering
    \includegraphics[width=0.65\linewidth]{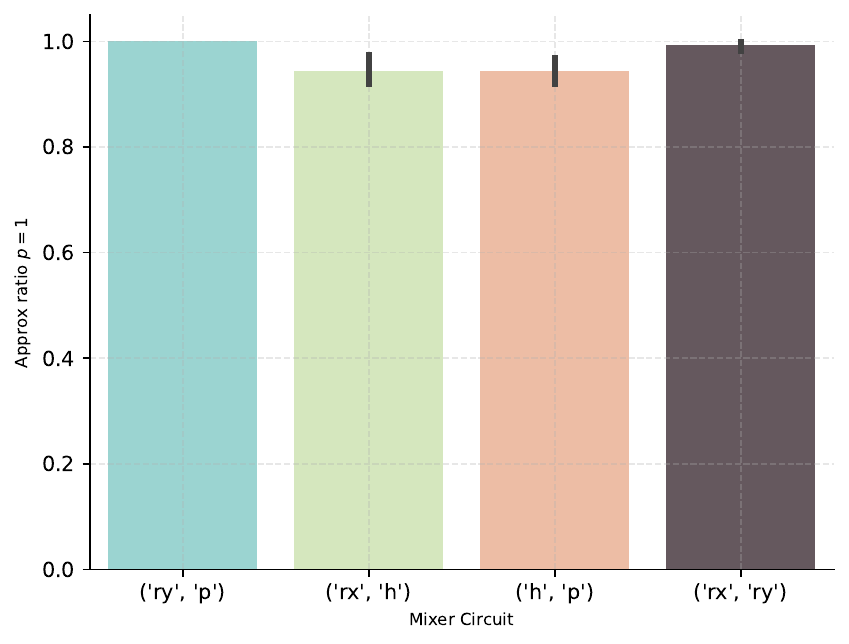}
    \caption{Approximation ratios obtained for $p=1$ on 
    4-regular random graphs. All parameterized gates in the mixer circuit share the same parameter and hence do not incur additional computational cost.}
    \label{fig:mixer_ckt_evaluation}
\end{figure}

Where $\langle C_{max} \rangle$ is the expected energy of the largest cut discovered by the given quantum circuit. The approximation ratio measures the quality of solutions discovered by the quantum procedure as compared to a classical one. The results are shown in Figure~\ref{fig:mixer_ckt_evaluation}. We can clearly see that the best performing mixer layer combination achieves the highest approximation ratio for low $p$ value.

\begin{figure}
    \centering
    \includegraphics[width=.65\linewidth]{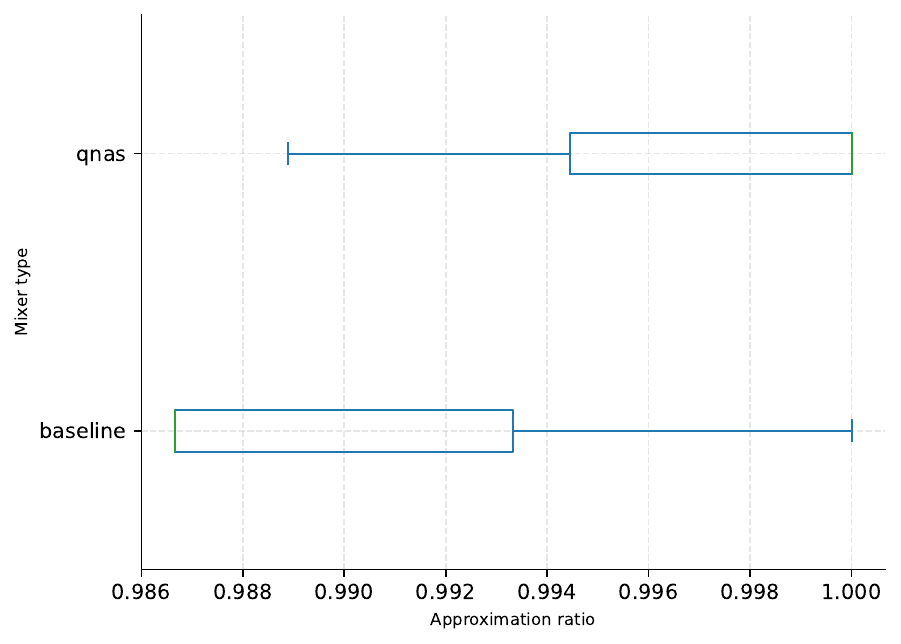}
    \caption{Comparison of $r$ obtained by the baseline and searched (qnas) mixer circuits}
    \label{fig:er_graph_comp}
\end{figure}

We further compare the performance of the searched mixer circuit on the ER and random regular graphs with the default mixer choice for maxcut QAOA. Figure~\ref{fig:er_graph_comp} shows the results for average $r$ obtained by the searched mixer and baseline mixer. These results were averaged by computing energies with $p = 1, 2, 3$. We can see that the searched mixer yields a higher average approximation ratio on ER random graphs. 

\begin{figure}
    \centering
    \includegraphics[width=0.65\linewidth]{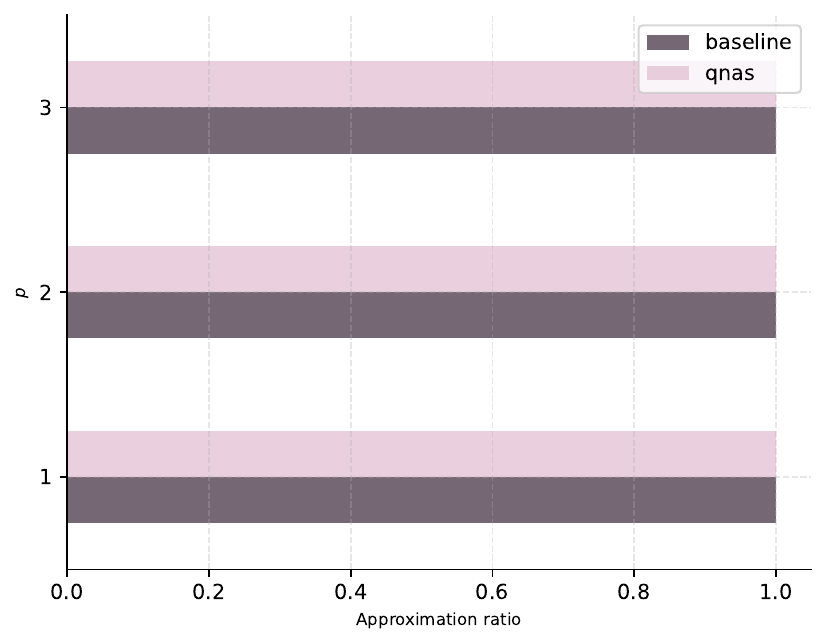}
    \caption{Approximation ratios obtained by baseline and qnas mixer circuits on 10 node random regular graph of degree 4.}
    \label{fig:random_graph_comp}
\end{figure}

In the case of random regular graph both mixer circuits perform comparably at all values of $p$. These results are shown in Figure~\ref{fig:random_graph_comp}. We show individual $r$ since the aggregated values over $p$ are equal ($1.0$). 

Overall, the mixer found by our search algorithm generally performs better with lower resource usage on different types of random graphs. Moreover, we show that our algorithm is able to extract the best \emph{general} performing circuit given data and some evaluation metric.

\section{Key Challenges and Future Work}

In this paper we have demonstrated that parallelizing the search algorithm is extremely important and requires careful design to obtain a meaningful speedup for large problem instances. We now discuss some important directions that are currently in development for \texttt{QArchSearch}.\\

\textbf{GPU Integration}: In this work, we noted that simulating quantum circuits was another computational bottleneck. However, improving the run time of quantum circuit simulations requires a different strategy since state vectors cannot be arbitrarily chunked and passed to multiple CPUs in a cluster. One possible way to improve the runtime is then to consider running the simulations on a GPU device. The future versions of \texttt{QArchSearch} will tightly integrate with \texttt{QTensor} to allow a user to seamlessly select a GPU backend whenever possible.\\ 

\textbf{Deep Neural Network based Search}: In this work we employed a version of random search to search for possible combinations of mixer circuits for the maxcut QAOA problem. Since this was a less complex problem than searching for full quantum circuits, random search returned strong generalized mixer circuits. However, our aim is to discover best quantum circuits for \emph{any} given dataset  and performance measure. \\ 

In the upcoming version of \texttt{QArchSearch} we will integrate several deep neural network based search algorithms like~\citep{zoph2016neural, zhou2018resource}. 

\section{Related Work}

Quantum architecture search is a very important and active area of research and different works have considered the problem from different angles.

Fosel~\emph{et al}~\citep{fosel2021quantum} consider the problem of optimizing the design of quantum circuits by first proposing inefficient circuits and then training a DNN to optimize the circuit given a desired circuit metric (e.g. number of gates). Ostaszewski~\emph{et al}~\citep{ostaszewski2021reinforcement} also propose to use deep reinforcement learning (RL) to obtain a good circuit for solving VQE~\citep{VQE} problem. Another hybrid method is considered by~\citep{duong2022quantum} where they propose to use Bayesian Optimization (BO) to discover optimal circuit architectures for a QNN given a particular dataset and loss function. Finally, a pure quantum architecture search is proposed by Du~\emph{et al}~\citep{du2022quantum} where they utilize a quantum ``supercircuit" to search for child quantum circuits that satisfy a given metric. To reduce computational cost, the parameters are shared amongst all child circuits.\\ 

In the hybrid approach (i.e. using DNN to discover circuits) a major bottleneck is the sample-inefficient nature of RL algorithms. Typically, it takes days if not weeks to find good candidate architecture on a given dataset. We note that our proposed software also falls in the hybrid category of algorithms. Our objective with \texttt{QArchSearch} is to reduce this search time to a couple of hours.One of the reasons we do not opt for a pure quantum architecture search procedure is due to the inherent issues of scalability. For instance~\citep{du2022quantum} note that they are unable to search for circuits beyond 2 or 3 qubits. In order for a general architecture search package to be useful, we desire that it scale to arbitrarly as many qubits as desired.

\section{Conclusions}

In this work, we demonstrated the implementation of \texttt{QArchSearch} package that finds short-depth compact quantum circuits and architectures for a given objective function using quantum simulator \texttt{QTensor}. \texttt{QArchSearch} runs at scale and high efficiency on high-performance computing systems using two-level parallelization scheme on both CPUs and GPUs, which has been demonstrated on the 44-Petaflop supercomputer Polaris located in Argonne Leadership Computing Facility~\citep{Polaris}.

Our software satisfies a critical need in the quantum computing community - a scalable software that automates searching of candidate quantum architectures for a given problem. Our software can also incorporate arbitrary constraints in the search procedure and thus deliver custom architectures that exceed performance of manually designed ones. In order to satisfy the needs of scalability and speed, we leverage reinforcement learning techniques running on GPUs and parallelize the search process on a large scale HPC system. Our belief is that our software will enable quantum computing researchers to find shorter-depth circuits for various advanced applications in the field.

\section{Acknowledgements}

 This work used in part the resources of the Argonne Leadership Computing Facility, which is a Department of Energy Office of Science User Facility supported under Contract DE-AC02-06CH11357. The views, opinions and/or findings expressed are those of the authors and should not be interpreted as representing the official views or policies of the Department of Energy or the U.S. Government. This work was supported  in part with funding from the Defense Advanced Research Projects Agency (DARPA).

\bibliographystyle{ACM-Reference-Format}
\bibliography{ref}







%



\end{document}